\documentclass[aps,prl,twocolumn,showpacs,amsmath]{revtex4}

\usepackage{dcolumn}% Align table columns on decimal point
\usepackage{bm}% bold math
\usepackage{amsfonts}
\usepackage{graphicx,graphics}
\usepackage{amssymb}
\usepackage{rotate}
\usepackage{subfigure}
\usepackage{epsf}
\usepackage{epsfig}
\usepackage{amsmath}
\usepackage{dcolumn}
\usepackage{color}
\usepackage{tikz}
\usepackage{hyperref}

\renewcommand{\=}{\,=\,}

\def \t {tetrahedron }

\begin{document}

\title{Spin ice under pressure: symmetry enhancement and infinite  order multicriticality}

\author{L.~D.~C.~Jaubert,$^{1}$ J. T. Chalker,$^{2}$ P. C. W. Holdsworth,$^{3}$ and R. Moessner$^{1}$}

\affiliation{$^{1}$Max-Planck-Institut f\"ur Physik komplexer Systeme, 01187 Dresden, Germany.} 
\affiliation{$^{2}$Theoretical Physics, Oxford University, 1 Keble Road, Oxford, OX1 3NP, United Kingdom.}
\affiliation{$^{3}$Laboratoire de Physique, \'Ecole Normale Sup\'erieure de Lyon, Universit\'e de Lyon, CNRS, 46 All\'ee d'Italie, 69364 Lyon Cedex 07, France.}

\date{\today}

\begin{abstract}
We study the low-temperature behaviour of spin ice when uniaxial pressure induces a tetragonal distortion. There is a phase transition between a Coulomb liquid and a fully  magnetised phase. Unusually, it combines features of discontinuous and continuous transitions: the order parameter exhibits a jump, but this is accompanied by a divergent susceptibility and vanishing domain wall tension. All these aspects can be understood as a consequence of an emergent SU(2) symmetry at the critical point. We map out a possible experimental realisation.
\end{abstract}


\maketitle

One fascinating aspect of condensed matter physics is the extent to which the nature and symmetries of emergent low-energy degrees of freedom can be independent of the high energy ones from which they derive. While many types of low energy degrees of freedom are possible,  systems in which these correspond to a gauge field are still rare.

In this paper we study this theme in a setting which brings forth an unusual sequence of low energy fluctuations and symmetries: we consider spin ice, a frustrated Ising magnet on the pyrochlore lattice, whose low temperature behaviour is well-described by a gauge field, an emergent magnetostatics, representing a so-called Coulomb phase where correlations decay algebraically as $r^{-3}$. Transitions out of such a phase have attracted a great deal of attention recently~\cite{Henley09,Alet06a,Pickles08a,Chen09}.

We ask what happens when the host crystal is subjected to uniaxial pressure. This is a powerful probe of correlated matter, which has previously produced interesting information in frustrated magnets closely related to spin ice~\cite{Mirebeau02a,Mirebeau06a}. Specifically, 
we consider the case of pressure-induced strain that lowers the crystal symmetry from cubic to tetragonal. We find that this induces a highly unusual symmetry-breaking phase transition out of the Coulomb phase: it is characterised by discontinuities in quantities such as the magnetisation, without being first order. Rather, it occurs at an infinite order multicritical point and is accompanied by a divergent susceptibility. We demonstrate both these features using Monte Carlo simulations.  Further salient properties of the critical point include the absence of a domain wall tension, along with spin correlations that vanish in a plane perpendicular to the strain-induced tetragonal axis. These should be observable experimentally, and we discuss the scope for realising such a transition in the laboratory.

The unusual nature of the transition is a consequence of an emergent SU(2) symmetry: as the pressure reduces the symmetry of the \textit{external} space, an enhanced {\em internal} symmetry appears. We show this by means of an exact solution in three dimensions at the critical temperature and a mapping to a quantum phase transition in $2+1$ dimensions, between Ising and XY anisotropy in a spin-$1/2$ ferromagnet. 

Transitions with analogous features have been studied previously in the context of ferroelectrics \cite{LinesGlass}, using versions of the model for potassium dihydrogen phosphate (KDP) introduced by Slater \cite{Slater41}, which in its two dimensional form is equivalent to the 6-vertex model and exactly solved \cite{Lieb67d}. Spin ice has the potential to provide a much cleaner realisation of this physics than the transition in ferroelectrics, since its magnetic degrees of freedom are much more accurately represented by Ising variables supported on a rigid lattice.

\begin{figure}[h,b]
\includegraphics[width=4cm]{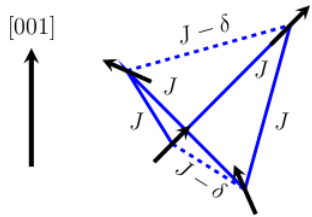}
\caption{A \t of the pyrochlore lattice with exchange $J$ and $J-\delta$ as indicated: the dashed bonds lie in (001) planes. One of the two ground states for $\delta > 0$ is shown.}
\label{fig:pyro}
\end{figure}

Spin ice is well modelled by Ising spins $\mathbf{S}_i$ parallel to their local easy-axis on the pyrochlore lattice~\cite{Harris97a}.
%(see Fig.~\ref{fig:pyro})
At sufficiently low temperature we can limit ourselves to the highly degenerate ground state ensemble with two spins pointing in and two out of each tetrahedron (ice-rules) and consider effective nearest neighbour interactions~\cite{Gingras01a,Isakov05a}. After coarse-graining the magnetisation, the discrete ice-rules can be written in terms of a continuous divergence free condition from which emerge the algebraic correlations~\cite{Isakov04b,Henley05a}. Our goal is to lift the degeneracy of the resulting Coulomb phase via the exchange modulation
\begin{eqnarray}
\mathcal{H}\=-\sum_{\left<i,j\right>}\,J_{ij}\,\mathbf{S}_{i} \cdot \mathbf{S}_{j}  
\label{eq:hamiltonian}
\end{eqnarray}
where $J_{ij}=J-\delta>0$ for bonds on $(001)$ planes perpendicular to the strain axis, and $J_{ij}=J$ otherwise (see Fig.~\ref{fig:pyro}). For $\delta>0$, the $\mathbb{Z}_{2}$ symmetry of this Hamiltonian is spontaneously broken at temperature $T_c\sim \delta$ in favour of a state magnetised parallel to the [001] direction. If the strain were very large, so that $T_c \sim J$, ordering would be from a conventional paramagnetic state and in the standard three-dimensional Ising universality class. In contrast, for realistic, small strains $T_c$ is much less than $J$ and ordering is from the Coulomb phase, with the striking features we describe.
\vspace{-0.2cm}

\begin{figure}[h]
\includegraphics[width=8.5cm]{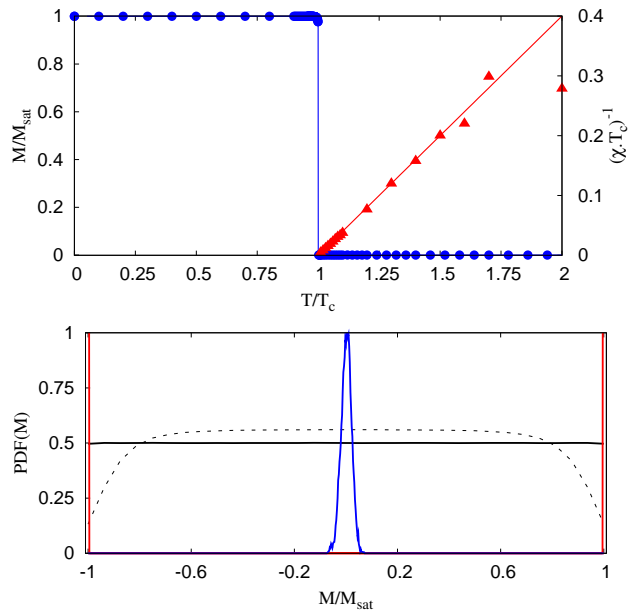}
\vspace{-0.1cm}
\caption{(Color online) \textit{Top:} Simulation data for $M$ (blue circles) and $(\chi T_c)^{-1}$ (red triangles) vs. $T/T_{c}$. \textit{Bottom:} PDF of  $M/M_{sat}$ at different temperatures. For $T<T_{c}$ (red) the PDF has two sharp peaks at $M=\pm M_{sat}$. For $T=T_{c}$ (black) the PDF is broad in small samples (dashed black line for $L_{\perp}=L_{z}=4$) and almost flat in larger samples (solid black line for $L_{\perp}=4$ and $L_{z}=18$). For $T>T_{c}$ (blue) the PDF is Gaussian. (Values at $T\not= T_c$ scaled to fit on vertical axis.)}
\label{fig:M}
\end{figure}

We first present results from Monte Carlo simulations,  done with a worm algorithm\cite{Isakov04a,Jaubert09c} in which loop updates preserve the divergence free nature of the magnetisation. The temperature dependence of the magnetisation $M$ and of the inverse susceptibility $\chi^{-1}$ are shown in Fig.~\ref{fig:M}, upper panel: remarkably, $M$ takes its saturation value $M_{sat}$ at all temperatures below $T_c$, but $\chi$ diverges as $T_c$ is approached from above. To better characterise the transition, we also examine the probability distribution function (PDF) of the order parameter, shown in Fig.~\ref{fig:M}, lower panel. This is a standard diagnostic: two peaks are expected in the PDF at the transition point when this is first order, but only one peak if it is continuous \cite{Binder87}. Again, we find unconventional behaviour: a PDF that is uniform \textit{over all values} of $M$ at $T_c$. 

An exactly uniform order parameter distribution arises within Landau theory as a limiting case.
The free energy density $G(M)$ close to a multicritical point of order $n$ is
\begin{eqnarray}
G(M)=\frac{\alpha_{2}}{2}(T-T_{c})M^{2}+\sum_{n'=n}^{+\infty}\frac{\alpha_{2n'}}{2n'}M^{2n'}\,.
\label{eq:gibbs}
\end{eqnarray}
In the limit $n\rightarrow \infty$, it is independent of $M$ at $T=T_c$. Moreover, the order parameter exponent takes the value $\beta=1/(2n-2)$, generating a jump for $n\to \infty$. Also, the susceptibility ($\chi \sim A_{\pm} |(T-T_c)/T_c|^{-\gamma}$ for $T\gtrless T_c$) has an exponent $\gamma=1$ independent of $n$ and an amplitude ratio $A_{-}/A_{+} = 1/(2n-2)$ that vanishes as $n\to \infty$. All these are borne out by our simulations. This success of mean field theory
reflects the exceptional value of the upper critical dimension,
$d_{c2}=1$, which follows from standard arguments with allowance
for the anisotropic spatial scaling,  see Eq~\ref{eq:corr} below.

In previous discussions of the ferroelectric KDP model~\cite{LinesGlass,Slater41,Takahashi41,Takagi48b,Nagle69a,Shore94a} 
%and of its application to crystal shapes~\cite{Shore94a} 
some distinctive aspects have been recognised. Indeed, Slater's original approximation \cite{Slater41} yields a free energy independent of the order parameter at the transition, which on this basis has been called infinite order \cite{Benguigui77a}. Such a conclusion is evidently very delicate \cite{Takahashi41},  and corrections to the approximation or physical perturbations have the potential to convert the transition into a conventional one: either first order or continuous. This is in fact the case for the material KDP itself, which has a first order transition that can be driven through a tricritical point under pressure \cite{LinesGlass}.
While in two dimensions, an exact solution of the KDP model corroborates the main features of Slater's results~\cite{Lieb67d,Shore94a,Baxter07a}, it is not a priori clear -- not least in view of the famous difficulties in generalising Onsager's solution of the Ising model -- what this tells us for higher dimension. Here we present simulations and exact results for the transition in three dimensions. In addition, we show how to realise it in a magnetic material.
%Conversely, for the KDP model the main features of Slater's results are corroborated by an exact solution in two dimensions \cite{Lieb67d,Shore94a,Baxter07a}. Here we present  simulations and exact results for the transition in three dimensions and show how to realise it in a magnetic material.

\vspace{-0.2cm}

The ice rules impose the same value of the magnetisation $M$ in all $(001)$ planes: any configuration can be mapped onto an ensemble of strings of down spins spanning the system from top to bottom, with the convention that the string vacuum is the state in which all spins are up \cite{Takahashi41}. The number of strings defines $(i)$ the total magnetisation and $(ii)$ a topological sector, so that the former can be used to label the latter. The equiprobability of all sectors  (see Fig.~\ref{fig:M}) already suggests the \textit{absence of interactions between strings}. This observation can be made precise using the transfer matrix that acts between two adjacent (001) layers of the lattice \cite{Powell08a}. Since this transfer matrix is a direct product of factors $\bf T$ representing separate tetrahedra, we first discuss a single tetrahedron. 
A single string can enter the tetrahedron at either site in one $(001)$ layer and leave it at either site in the other layer, with energy cost $4\delta/3$. This sector is therefore represented in $\bf T$ by a  $2\times2$ block in which all entries are the Boltzmann factor $\Gamma=e^{-4\delta\beta/3}$.  By contrast, zero or two strings in a tetrahedron cost no energy and impose the next configuration locally, generating two $1\times 1$ blocks with unit entries in $\bf T$, which hence has the form
\begin{equation}
{\bf T} = \left(
\begin{array}{cccc}
1 & & & \\
& \Gamma &\Gamma & \\
& \Gamma &\Gamma & \\
& & & 1
\end{array}
\right)\,.
\end{equation}
The maximum eigenvalues of $\bf T$ lie for $T<T_c=\frac{4\delta}{3\ln 2}$ in the two fully magnetised sectors, but for $T=T_c$ all sectors have the same maximal eigenvalue. In addition, the associated eigenvector in the one-string sector is $(1,1)$. Explicit calculation of the transfer matrix for a complete lattice is complicated by the fact that the repeat unit in the $[001]$ direction involves four different layers. Nevertheless, its maximal eigenvalues inherit precisely the properties described for $\bf T$, and at $T=T_c$ the associated eigenvectors in each sector give equal weight to all arrangements of strings on a $(001)$ layer. The maximal eigenvalues determine the physical properties for a system of size $L_z\times L_\perp^2$ if the $[001]$ dimension satisfies $L_z\gg L_\perp^2$. As a result: (i) below $T_c$ {\em the magnetisation is saturated;} and (ii) at $T_c$ {\em all sectors are equiprobable} and {\em all configurations within a sector are equiprobable}, reflecting the absence of interactions between strings~\cite{Jaubert09c}. By contrast, for a system with $L_z\sim L_\perp$ subleading eigenvalues also contribute and the PDF for $M$ is rounded, as is apparent in Fig.~\ref{fig:M}. 

The transfer matrix for the complete lattice can alternatively be thought of as the evolution operator in imaginary time for a two-dimensional quantum $XXZ$ model (in analogy with Eq. (8) of \cite{Lieb67d}) with Hamiltonian
\begin{eqnarray}
\mathcal{H}_Q=-J\sum_{\left<i,j\right>}\left(s_{i}^{x}s_{j}^{x}+s_{i}^{y}s_{j}^{y}+\Delta\,s_{i}^{z}s_{j}^{z}\right)\,.
\label{eq:ham1}
\end{eqnarray}
The strings thus denote the world lines of $s^z_i = 1/2$ spins. In this language the decomposition into sectors is a consequence of the conservation of total $s^z$, and is due to a U(1) symmetry.
The ground state of $\mathcal{H}$ has ferromagnetic order of the quantum spins for all $\Delta$, with an orientation for the magnetisation that depends on $\Delta$. For $\Delta<1$, spins lie in the $x$-$y$ plane and the state has zero-point fluctuations, representing the Coulomb phase. For $\Delta>1$, spins are aligned along the $z$-axis and the state has no fluctuations, representing the  low temperature phase of the classical system. The equiprobability of string sectors at the critical point corresponds here to the degeneracy of the $N+1$ ground states of the isotropic Heisenberg ferromagnet with $N$ spins, for all values of the total magnetisation $\Sigma^{z}=\{-N/2,..,N/2\}$. We have confirmed that this isotropy is a property of the full transfer matrix, and not only of the leading eigenvectors, by checking that the matrix commutes with the total spin raising and lowering operators, $\Sigma^{\pm} = \sum_j(s_j^x \pm i s_j^y)$. This constitutes an enhancement of the symmetry at the critical point from U(1) to SU(2).

The quantum description can be employed to calculate correlation functions within a given sector in spin ice, by approximating the imaginary time direction as continuous and treating ${\cal H}_Q$ using harmonic spinwave theory. For a system of shape sufficiently anisotropic ($L_z \gg L_\perp^2$) to justify a ground state treatment of the 2+1 dimensional problem, and taking $z$ and $\mathbf{r}$ to denote distances along the [001] direction and in the (001) plane respectively, with $\rho$ a microscopic length, two-point correlations at $T_c$ are
\begin{eqnarray}
C(\mathbf{r},z)\propto \frac{1}{z}\;\exp\left(-\frac{|\mathbf{r}|^{2}}{\rho\,z}\right)\,.
\label{eq:corr}
\end{eqnarray}
As the strings do not interact, this form reflects the string auto-correlations present in a random walk in two dimensions with propagation time $z$. It agrees well with results from simulations, as we show in Fig.~\ref{fig:corr}. 
% and is in good agreement with numerical simulations, even in the natural scaling limit $L_{z}\sim L_{\perp}^{2}$, up to finite size corrections (see Fig.~\ref{fig:corr}). This strong anisotropy suggests two different correlation length exponents $\nu_{\parallel}=1$ and $\nu_{\perp}=1/2$~\cite{Bhattacharjee91b}, giving an upper-critical dimension $d_{c}=1$ instead of 2 for an isotropic infinitely multi-critical point~\cite{Benguigui77a}.
%
\definecolor{grey}{rgb}{0.525,0.525,0.525}
\definecolor{pink}{rgb}{1,0,1}
\begin{figure}[t]
\includegraphics[width=8cm]{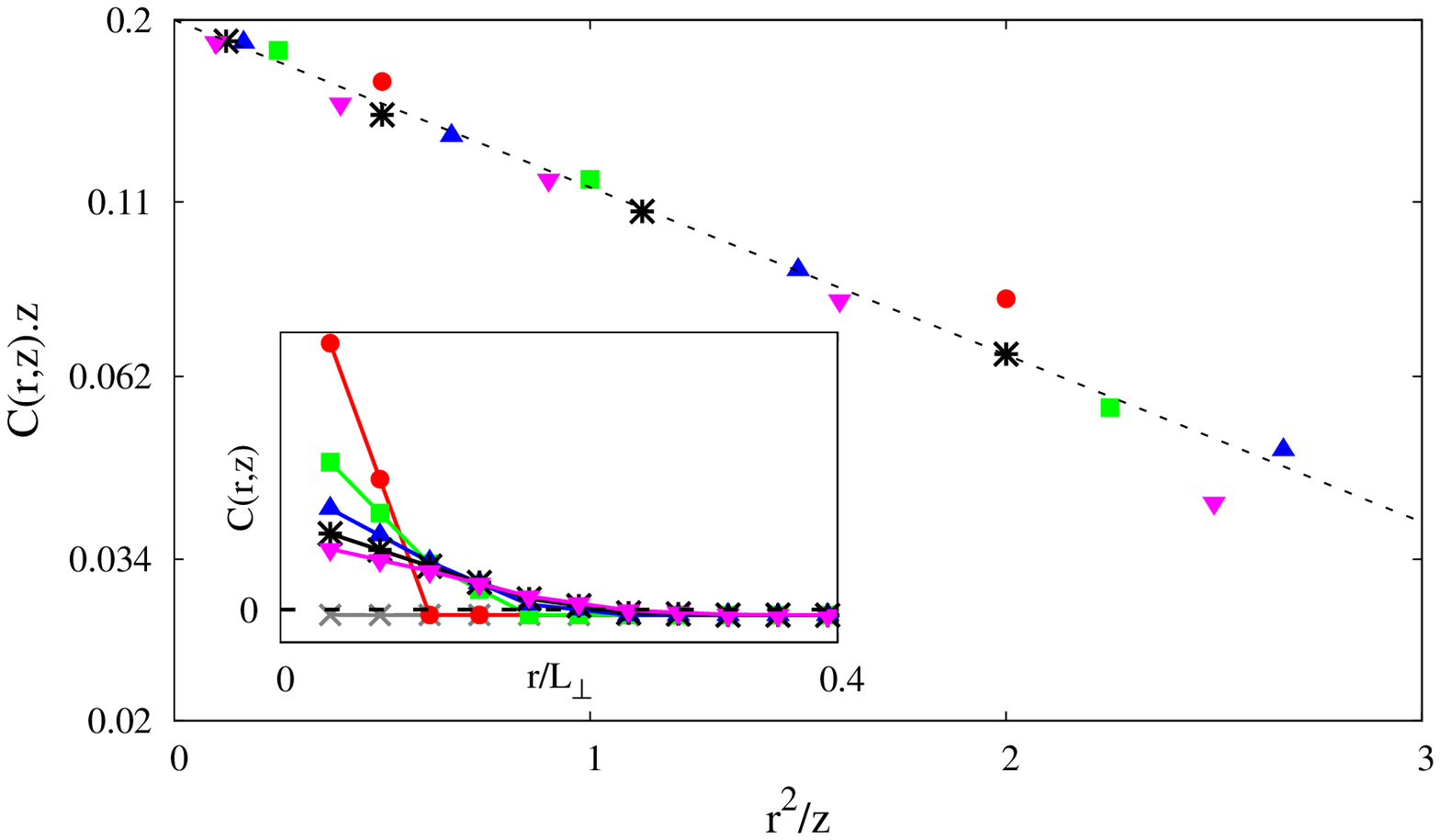}
\vspace{-0.2cm}
\caption{\textit{(Color online)} \textit{Main panel:} Behaviour of $z\cdot C(\mathbf{r},z)$ vs $r^{2}/z$, on a semi-log scale for $z=0$ (\textcolor{grey}{$\times$}), 2 (\textcolor{red}{$\bullet$}), 4 (\textcolor{green}{$\blacksquare$}), 6 (\textcolor{blue}{$\blacktriangle$}), 8 (\textcolor{black}{$\mathbf{+}$}), 10 (\textcolor{pink}{$\blacktriangledown$}). Dashed line shows behaviour expected from Eq.~(\ref{eq:corr}). The system size is $L_{\perp}=14$ and $L_{z}=196$. \textit{Inset:} Same data plotted as $C(\mathbf{r},z)$ vs $r/L_{\perp}$ on a linear scale, showing that correlations vanish within (001) planes.}
\label{fig:corr}
\end{figure}

\begin{figure}[h]
\includegraphics[width=8.5cm]{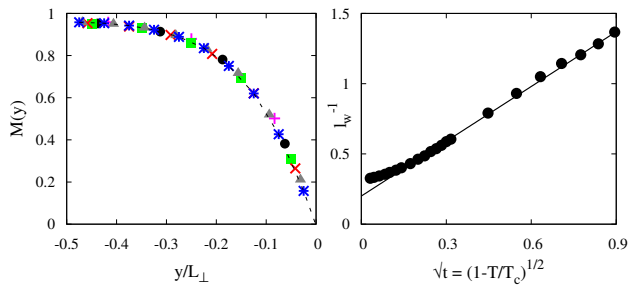}
\vspace{-0.2cm}
\caption{\textit{Left:} Magnetisation profile vs $y/L_{\perp}$ at $T=0.999\,T_{c}$ for $L_{\perp}=\{6,8,10,12,16,20\}$. \textit{Right:} Domain wall width $\ell_{w}$ vs $\sqrt{t}$ for $L_{\perp}=20$. The straight line is  a guide to the eye.
% its intercept at $t=0$ scales to zero with increasing system size. 
All samples have $L_{z}=L_{\perp}^{2}$.}
\label{fig:domwall}
\end{figure}

A further consequence of the absence of interactions between strings is that the surface tension between oppositely magnetised domains vanishes and the domain wall width diverges, as $T_c$ is approached from below. We investigate this phenomenon in simulations by using weak, position-dependent magnetic fields to induce two domains with the interfaces between them lying on average in the $x$-$z$ plane. Results are shown in Fig.~\ref{fig:domwall}.
At $T_{c}^{-}$ the magnetisation profile $M(y)$ is a function only of $y/L_\perp$. Below $T_c$ a mean field treatment of ${\cal H}_Q$ gives a domain wall width 
$\ell_{w}^{-1}\propto \sqrt{t}\equiv(1-T/T_{c})^{1/2}$: this is confirmed in Fig.~\ref{fig:domwall}, up to a constant vanishing in the thermodynamic limit. It is striking to find broad domain walls in an Ising magnet; they may be detectable using small angle neutron scattering.
The natural way to generate the required degeneracy lifting in a magnetic compound is to apply uniaxial pressure along the [001] axis of a single crystal. We require $\delta$ positive. Within the nearest neighbour approximation the effective coupling has two contributions, one from the long-range dipolar interactions~\cite{Siddharthan99a,Gingras01a,Isakov05a}, the other due to superexchange.
While a uniaxial compression along [001] increases the former, the change of the latter is less clear as it depends on the evolution of orbital overlaps. 
\begin{figure}[h]
\includegraphics[width=8.cm]{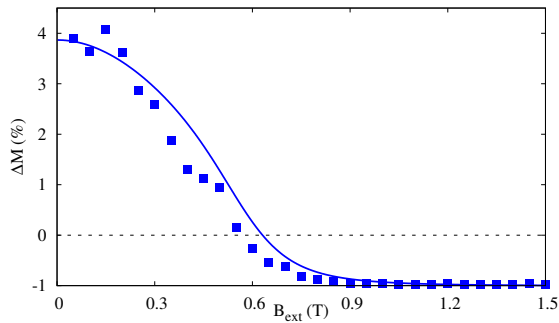}
\caption{Percentage change in magnetisation measured under uniaxial pressure of $10.5$ kbar, $M(P,B)$, compared to that at atmospheric pressure, $M(0,B)$,  at $T=1.7$K as a function of applied field $B$ along the $[001]$ axis (extracted from~\cite{Mito07a}).}
\label{fig:mito}
\end{figure}

Results from the one experiment so far performed on Dy$_{2}$Ti$_{2}$O$_{7}$ under pressure~\cite{Mito07a} suggest that the effective coupling is indeed modified so as to produce $\delta>0$. Pressure causes a $4\%$ increase in $M$ at small field, but a $1\%$ decrease at high field (Fig.~\ref{fig:mito}). The increase in the  zero field susceptibility is consistent with the lifting of the degeneracy in favour of the states with magnetization along the field axis, as required for the transition studied here (Fig.~\ref{fig:pyro}). The reduction of the saturated moment is expected as the crystal is squeezed along [001] because of a resulting tilt of the easy axes away from [001]. 

In fitting  the parameter $\delta$, we account for demagnetisation effects by modelling the  sample  as a prolate ellipsoid with major axes given by the dimensions of the approximately rectangular parallelepiped specimen.  For the interaction parameter of Dy$_{2}$Ti$_{2}$O$_{7}$~\cite{Hertog00a}, the best fit, shown in Fig.~\ref{fig:mito}, yields $T_{c}\sim 200$ mK.

Whereas this temperature is comfortably within reach of cryogenics, the dynamics of spin ice slows down greatly below $T_f \sim 600$mK, as activated ice-rule violating defects disappear. Their presence thus appears indispensable in practice, even though they will lead to more standard second-order phase transition in the three dimensional Ising universality class~\cite{Takagi48b,Jaubert09c}. The width of its critical region will, however, vanish with defect concentration. Experiments should thus aim at the lowest dynamically accessible transition temperatures, with the target $T_c \sim 600$mK only a factor of 3 away from existing experiments~\cite{Mito07a}. We hope our work will stimulate further experimental efforts to realise this unusual transition.

We thank K. Matsuhira for sharing results from Ref.~\cite{Mito07a}, and D. Simon and S.T. Bramwell for useful discussions. We acknowledge financial support from ESF under Grant PESC/RNP/HFM (PCWH and LJ), from ANR under Grant 05-BLAN-0105 (LJ), and from EPSRC (JTC) under Grant EP/D050952/1.

\end{document}